\documentclass[12pt,oneside]{amsart}
\usepackage{amsmath}               
\usepackage{amsthm}                
\usepackage{times}
\usepackage{hyperref}
\theoremstyle{plain}\newtheorem{teo}{Theorem}[section]
\theoremstyle{plain}\newtheorem{lem}[teo]{Lemma}
\theoremstyle{plain}\newtheorem{prop}[teo]{Proposition}
\theoremstyle{plain}
\theoremstyle{definition}
\theoremstyle{definition}\newtheorem{assum}[teo]{Assumption}
\theoremstyle{remark}\newtheorem{rem}[teo]{Remark}
\theoremstyle{remark}
\numberwithin{equation}{section}
\title[Naked singularities in perfect fluids gravitational collapse]%
{Naked singularities formation in the gravitational collapse of
barotropic spherical fluids}
\author[R.\ Giamb\`o ,\ F.\ Giannoni]{Roberto Giamb\`o, Fabio Giannoni}
\address{Dipartimento di Matematica e Informatica,
Universit\`a di Camerino, Italy} \email{roberto.giambo@unicam.it,
fabio.giannoni@unicam.it}
\author[G.\ Magli]{Giulio Magli}
\address{Dipartimento di Matematica, Politecnico di
Milano, Italy} \email{magli@mate.polimi.it}
\author[P.\ Piccione]{Paolo Piccione}
 \address{Dipartimento di Matematica e Informatica,
Universit\`a di Camerino, Italy \hfill\break\indent (on leave
from Departamento de Matem\'atica,\hfill\break\indent
Universidade de S\~ao Paulo, Brazil)}
\email{paolo.piccione@unicam.it, (piccione@ime.usp.br)}
\begin{document}
\begin{abstract}
The gravitational collapse of spherical, barotropic perfect fluids
is analyzed here. For the first time, the final state of these
systems is studied without resorting to simplifying assumptions -
such as self-similarity - using a new approach based on non-linear
o.d.e. techniques, and formation of naked singularities is shown
to occur for solutions such that  the mass function is
analytic in a neighborhood of the spacetime
singularity.
\end{abstract}
\maketitle
\section{Introduction}
The final state of gravitational collapse is an open problem of
classical gravity. It is, in fact, commonly believed that a
collapsing star that it is unable to radiate away - via e.g.
supernova explosion - a sufficient amount of mass to fall below
the neutron star limit, will certainly and inevitably form a black
hole, so that the singularity corresponding to diverging values of
energy and stresses will be safely hidden - at least to faraway
observers - by an event horizon. However, this is nothing more
than a conjecture - what Roger Penrose first called a "Cosmic
Censorship" conjecture \cite{pen} - and has never been proved.
Actually, it is easy to see that one just cannot prove the
conjecture as a statement on the mathematical evolution of {\it
any} collapsing system via Einstein field equations, because in
this case what is conjectured is baldly false: it is indeed an
easy exercise producing counterexamples using e.g. negative energy
densities or "ad hoc" field configurations. Thus, to go beyond the
conjecture what is needed is a set of hypotheses, possibly based
on sound physical requirements, which would allow the proof of a
mathematically rigorous theorem. However, what turned out to be
the truth in the last twenty years of research is that such a
theorem (and, in fact, even the hypotheses of the theorem) is/are
extremely difficult to be stated (see e.g. \cite{jbook}).

In the meanwhile, many examples of spherically symmetric
solutions exhibiting naked singularities and satisfying the
principles of physical reasonableness have been discovered.

Spherically symmetric naked singularities can be divided into two
groups: those occurring in scalar fields models \cite{Ch94,Ch99}
and those occurring in astrophysical sources modeled with
continuous media, which are of exclusive interest here (see
\cite{jmc} for a recent review). The first (shell focusing)
examples of naked singularities where discovered in dust models,
numerically by Eardley and Smarr \cite{e2} and analytically by
Christodoulou \cite{e3}. Today, the gravitational collapse of dust
is known in full details \cite{sj}.

The dust models can, of course, be strongly criticized from the
physical point of view. In fact, they have the obvious drawback
that stresses are expected to develop during the collapse,
possibly influencing its dynamics. In particular, such models are
an unsound description of astrophysical sources in the late stage
of the collapse even if the latter does {\it not} form a
singularity: one can, for instance, regard a white dwarf or a
neutron star as being an extremely compact {\it planet}, composed
by a solid crust and a liquid (super)fluid core: such objects are
sustained by enormous amounts of (generally anisotropic) stresses.
It is, therefore, urgent to understand models of gravitational
collapse with stresses.

Recently, several new results have been obtained in this
direction by considering systems sustained by anisotropic
stresses (see e.g. \cite{gjm,ha,hacqg,r}). Besides the details of
the physics of the collapse of such systems, the general pattern
arising from all such examples is that existence of naked
singularities persists in presence of stresses: actually, we have
recently shown that the mechanism responsible for the formation
or whatsoever of a naked singularity is {\it the same} in all
such cases \cite{ns}.

 In spite of the aforementioned physical
relevance of anisotropic systems, it is beyond any doubt of
exceeding interest the case of {\it isotropic} stresses, i.e. the
gravitational collapse of perfect fluids. In fact, for instance,
the perfect fluid model is (in part for historical reasons) the
preferred model used in most approximations of stellar matter of
astrophysical interest. Unfortunately, although {\it local}
existence ad uniqueness for the solution of the Einstein field
equations has been proved \cite{ekind,rendall}, very few sound
analytical models of gravitational collapse of perfect fluids are
known and, as a consequence, the problem of the final state of
gravitational collapse of perfect fluids in General Relativity is
still essentially open. Exceptions are the solutions describing
shear-free fluids (see e.g. \cite{Kra,K}) and those obtained by
matching of shock waves \cite{smoller}; in both cases, however,
the collapse is synchronous (i.e. the singularity is of the
Friedmann-Robertson-Walker type) and therefore such solutions say
little about Cosmic Censorship \cite{bjm,jdad}.

There is a unique perfect fluid class of solutions which has been
investigated in full details. This is the case of self-similar
fluids, which has been treated by many authors since the
pioneering work by Ori and Piran \cite{opi} (for a recent review
see \cite{car}). Self-similarity is compatible with the field
equations if the equation of state is of the form $p=\alpha
\epsilon$ (where $p$ is the pressure, $\epsilon$ the energy
density, and $\alpha$ a constant). In this case the field
equations reduce to ordinary differential equations and therefore
can be analyzed with the powerful techniques of dynamical
systems. Ori and Piran found that self similar perfect fluids
generically form naked singularities; more precisely, they showed
numerically that for any $\alpha$ in a certain range there are
solutions with naked singularities. Recently, Harada added some
numerical examples which remove the similarity hypotheses
\cite{hare}.

These results clearly go in the direction of disproving {\it any}
kind of censorship at least in spherical symmetry, since they
show that naked singularities have to be expected in perfect
fluids with physically sound equations of state.  However,
although being extremely relevant as a "laboratory",  the
self-similar ansatz is a over-simplifying assumption, and the
general case of perfect fluid collapse remained untractable up
today, essentially due to the lack of exact solutions.

In the present paper we present the first (as far as we are aware)
analytical study on the endstates of barotropic spherical fluids
which circumvents this problem. To do this we use a combination of
two new ingredients. The first is the fact that, in a suitable
system of coordinates (the so-called area-radius coordinates) we
are able to reduce the field equations to a single, quasi linear,
second order partial differential equation. As a consequence, the
metric for a barotropic spherical fluid can be written, in full
generality, in terms of only one unknown function.  In this way
the behavior of the null radial geodesics near the singular point
can be analyzed in terms of the Taylor expansion of such a
function. The second ingredient is a new framework for doing this
analysis based on techniques for singular non linear ordinary
differential equations \cite{ns,new}.

Our results here show the existence of naked singularities in
barotropic perfect fluids solutions for which the mass function is
analytic in a neighborhood of the center.

\section{Reduction of the field equations to a quasi-linear P.D.E.}

Consider a spherically symmetric perfect fluid. The general line
element in comoving coordinates can be written as
\begin{equation}\label{eq:ds-tr}
\text ds^2=-e^{2\nu}\text dt^2 +e^{2\lambda}\text dr^2 +R^2 (\text
d\theta^2
+\sin^2\theta\, \text d\phi^2)
\end{equation}
where $\nu, \lambda$ and $R$ are function of $r$ and $t$ (we
shall use a dot and a prime to denote derivatives with respect to
$t$ and $r$ respectively). Denoting by  $\epsilon$ and $p$ the
energy density and the isotropic pressure of the fluid, Einstein
field equations can be written as
\begin{subequations}
\begin{align}
&\Psi' =4\pi \epsilon R^2 R',\label{ma1}\\
&\dot\Psi =-4\pi p R^2 \dot R,\label{ma2}\\
&\dot R'  = \dot R \nu' + R'\dot \lambda,\label{ma3}\\
&p'=-(\epsilon +p) \nu',\label{ma4}
\end{align}
\end{subequations}
where $\Psi(r,t)$ is the {\it Misner-Sharp mass}, defined in such
a way that  the equation $R=2\Psi$ spans the boundary of the {\it
trapped region}, i.e. the region in which outgoing null rays
re-converge:
\begin{equation}\label{ma}
 \Psi(r,t)=\frac R2\left[1-g^{\mu\nu}(\partial_\mu R )(\partial_\nu
R)\right]=
 \frac R2  \left[1-(R')^2e^{-2\lambda}+(\dot
R)^2e^{-2\nu}\right]\ ,
\end{equation}
The curve $t_h(r)$ describing this boundary, i.e. the function
defined implicitly by
\begin{equation}\label{eq:horizon}
R(r,t_{h}(r))=2\Psi(r,R(r,t_{h}(r))),
\end{equation}
is called  {\it apparent horizon} and will play a fundamental
role in what follows.

Initial data for the field equations can be assigned on any
Cauchy surface ($t=0$, say). Physically, the arbitrariness on the
data refers to the initial distribution of energy density and the
initial velocity profile, and is therefore described by two
functions of $r$ only. Data for $R$ do not carry physical
information and we parameterize the initial surface in such a way
that $R(r,0)=r$.

The data must be complemented with the information about the
physical nature of the collapsing material. In the present paper
we shall consider only barotropic perfect fluids, i.e. fluids for
which the equation of state can be given in the standard
thermodynamical form: the pressure $p$ equals minus the
derivative w.r. to the specific volume $v$ of the specific energy
density $e(v)$. We are going to work however with the matter
density $\rho=1/v$ and with the energy density $\epsilon
(\rho)=\rho e(1/\rho)$. Therefore we are going to use in the
sequel the equation of state of the fluid in the form (slightly
less familiar than $p=-de/dv$):
\begin{equation}\label{pe}
p=\rho \frac{d\epsilon}{d\rho}-\epsilon
\end{equation}
Using the comoving description of the fields the matter density
is proportional to the determinant of the 3-metric, i.e.
\begin{equation}\label{eq:rho} \rho
=\frac{e^{-\lambda}}{4\pi ER^2}
\end{equation}
where $E=E(r)$ is an arbitrary positive function.

In order to simplify reading, we are going to develop in full
details in the next sections the special - although physically
very relevant -  case of the linear equation of state
\begin{equation}\label{eq:T-bpf}
p=\alpha\,\epsilon,
\end{equation}
where $\alpha$ is a constant parameter. However, in the final
section, we will show how the results can be easily extended to
(virtually) {\it all} the - physically valid - barotropic
equations of state.

In terms of the matter density eq. \eqref{pe} implies
$\epsilon=\rho^{\alpha+1}$ up to a multiplicative constant which
however can be absorbed in the definition of $E(r)$. For such
fluids the field equation \eqref{ma4} integrates to
\begin{equation}\label{enu}
e^{\nu}=\rho^{-\alpha}
\end{equation}
up to a multiplicative function of time only which can be taken
equal to one by a reparameterization of $t$.

We are now going to show that the remaining field equations
simplify considerably (and actually the problem of the final
state becomes tractable) if another system of coordinates, the
area-radius ones, are used. The advantages of this system were
first recognized by Ori \cite{Ori}, who used it to obtain the
general exact solution for charged dust. Subsequently, the
area-radius framework has been successfully applied to models of
gravitational collapse and cosmic censorship (see e.g.
\cite{ns,ha,m2}).

Area-radius coordinates are obtained using $R$ in place of the
comoving time. Denoting by subscripts derivatives w.r. to the new
coordinates, we have $\Psi'=\Psi_{,r}+R'\,\Psi_{,R}$, $\dot
\Psi=\dot R\,\Psi_{,R}$. Substituting in eqs \eqref{ma1},
\eqref{ma2} we obtain $R'$ and $\rho$ in terms of the mass:
\begin{equation}\label{eq:Rprime}
R'=-\frac{\alpha}{\alpha+1}\frac{\Psi_{,r}}{\Psi_{,R}}.
\end{equation}
\begin{equation}\label{eq:efe1b}
\rho=\left(-\frac{\Psi_{,R}}{4\pi\alpha
R^2}\right)^{\frac{1}{\alpha+1}},
\end{equation}
In writing the above formulae we have excluded the case
$\alpha=0$. This case corresponds to the dust (Tolman-Bondi)
solutions which is already very well known and will not be
considered further in the present paper (see \cite{sj} and
references therein).

Equation \eqref{ma} can be used to express the velocity $u=|\dot
R e^{-\nu}|$ as
\begin{equation}\label{eq:efe2b}
u^2=\frac{2\Psi}{R}+Y^2-1.
\end{equation}
where we have introduced the function
\begin{equation}\label{eq:Ydef}
Y=R'\,e^{-\lambda},
\end{equation}
using \eqref{enu}, \eqref{eq:Rprime} and \eqref{eq:efe1b} we have
\begin{equation}\label{eq:Y}
Y=\frac{E\,\psi_{,r}}{(\alpha+1)\rho^\alpha}.
\end{equation}
This function plays the role of an ``effective potential'' for the
motion of the shells. Notice that $u$ is known when $Y$ and $\Psi$
are; $Y$ is known when $E(r)$ is given and $\Psi$ is known. Thus,
in particular, the initial velocity profile $u(r,r)$ is known when
the functions
\begin{equation}
\Psi_0 (r)=\Psi(r,r),\qquad Y_0(r)=Y(r,r)
\end{equation}
are
known. It is therefore convenient to use $Y_0$ as the second
arbitrary function, eliminating $E$:
\begin{equation}\label{eq:Ynew}
Y(r,R)=\frac{\Psi_{,r}(r,R)}{\Psi_{,r}(r,r)}
\left[\frac{\Psi_{,R}(r,r)\,R^2}{\Psi_{,R}(r,R)\,r^2}\right]^{\frac\alpha{
\alpha+1}}Y_0(r),
\end{equation}
where \eqref{eq:efe1b} and \eqref{eq:Y} have  been used.

We conclude that the metric for a barotropic perfect fluid in
area-radius coordinates can be written in terms of the data and
of the function $\Psi$ and its first derivatives  as follows:

\begin{multline}\label{met}
\text ds^2=-\frac 1{u^2} \left[\text dR^2-2R' \text dR\, \text dr
+ \left( \frac{R'}{Y}\right)^2 (1-\frac{2\Psi}R)\,\text
dr^2\right] +\\ +R^2 (\text d\theta^2 +\sin^2\theta\, \text
d\phi^2)
\end{multline}
where $u$, $R'$ and $Y$ are given by formulae \eqref{eq:efe1b},
\eqref{eq:Rprime} and \eqref{eq:Ynew} above. By a tedious but
straightforward calculation the remaining field equation can be
re-arranged as a second order equation for $\Psi$. Remarkably
enough, this equation is quasi-linear. In fact, the following
holds true:

\begin{teo}\label{rem:PDE}
The Einstein field equations for a spherical barotropic fluid in
the coordinate system \eqref{met} are equivalent to the
following, second order PDE:
\begin{equation}\label{eq:2PDE}
a\,\Psi_{,RR}+ 2b\,\Psi_{,rR}+c\,\Psi_{,rr}=d,
\end{equation}
where $a,b,c,d$ are functions of $r,R,\Psi,\Psi_r,\Psi_R$ given
by:
\begin{subequations}
\begin{align}
&a=\frac{1}{(\alpha+1)\Psi_{,R}}\left[1-\alpha\left(\frac
Yu\right)^2\right],\label{eq:a}\\
&b=\left(\frac Yu\right)^2\frac1{\Psi_{,r}},\label{eq:b}\\
&c=-\frac{(\alpha+1)\Psi_{,R}}{\alpha\,\Psi_{,r}^2}\left(\frac
Yu\right)^2,\label{eq:c}
\end{align}
\begin{multline}
d=\frac1R\left[
-\frac{2\alpha}{\alpha+1}\left(1+\left(\frac Yu\right)^2\right)+
\frac{\alpha\Psi+ R \Psi_{,R}}{\alpha u^2 R}\,+\right.
\\\left.+\frac{(\alpha+1)\Psi_{,R}}{\alpha \Psi_{,r}}
\left(\frac{Y_0'}{Y_0}-\frac{1}{\alpha
+1}\frac{\Psi_0''}{\Psi_0'}-\frac{2\alpha}{(\alpha+1)r}\right)
\left(\frac Yu\right)^2
R\right].\label{eq:d}
\end{multline}
\end{subequations}
\end{teo}
\begin{rem}\label{rem:s1}
Equation \eqref{eq:2PDE} must be supplemented with a set of data
on the surface $R=r$ . Since
\begin{equation}\label{eq:discr}
ac-b^2=-\frac1{\alpha}\left(\frac Y{u\Psi_{,r}}\right)^2,
\end{equation}
the character of the equation is determined by the sign of
$\alpha$. In particular, the equation is hyperbolic for positive
pressures and elliptic for the negative ones (recall that
$\alpha=0$ is excluded). For physical reasons, however, we
consider here only the hyperbolic case (see next section). The
initial data for equation \eqref{eq:2PDE} are thus given, in
principle, by two functions. The value of $\Psi$ on the data
corresponds to the physical freedom of assigning the initial mass
distribution, while the first derivative can be calculated using
eq. \eqref{eq:Rprime} evaluated on the data. On $R=r$ one has
$R'=1$ and therefore:
\begin{equation}\label{constraint}
\Psi_{,R} (r,r)= -\frac{\alpha}{\alpha+1}\Psi_{,r} (r,r).
\end{equation}
\end{rem}

\begin{rem}\label{rem:rend}
A perfect fluid solution need not form a singularity: one can have
oscillating, regular spheres as well. This poses the problem of
characterizing the space of initial data w.r. to the final state
(regular or singular). As far as we know this problem has never
been studied (of course, it raises the issue of global existence
that, as known, is extremely difficult) so that results
like those known in the case of Einsten-Vlasov systems, for which
`small` (in a precise analytical sense) data lead to globally
regular solutions \cite{rr} are not available here. In what
follows, we are not going to address this problem. Therefore, we
proceed further considering those data that lead to singularity
formation with analytic mass function.
It is, at present, unclear the degree of genericity of
such data within the whole space of avaliable data, and this will
be the subject of future work.
\end{rem}

\begin{rem}\label{rem:s}
Equation \eqref{eq:2PDE} becomes degenerate at the sonic point,
when the relative velocity of the fluid equals the speed of sound.
The behavior of the solutions at the sonic point is quite
complicated, and not all the solutions can be extended. The
problem of characterizing the structure of the space of the
solutions is extremely interesting. As far as the present authors
are aware, such an analysis has been carried out in full details
only in the self-similar case \cite{bogo,opi,foglizzo}. In the
present paper, however, we are interested only in singularities
which arise from the gravitational interaction.
\end{rem}

\section{Formation and nature of singularities}

\subsection{Physical requirements}\label{subsec:phys}

We are going to impose here strict requirements of physical
reasonableness. First of all, we impose the dominant energy
condition, namely, energy density must be positive and the
modulus of the pressure cannot exceed the energy density (so that
$-1\leq \alpha \leq 1$). We consider, however, only the case of
positive pressure. It must, in fact, be taken into account that,
while tensions are common in anisotropic materials, a perfect
fluid can hardly be considered as physical in presence of a
negative isotropic pressure.

Therefore, $\alpha>0$ and \eqref{eq:efe1b} imply that
\begin{equation}\label{eq:PsiR}
\Psi_{,R}(r,R)<0,\qquad\forall r>0,\quad\forall R\in [0,r],
\end{equation}
and since we want $R'>0$ to avoid shell--crossing singularities (see below),
it must also be, from \eqref{eq:Rprime},
\begin{equation}\label{eq:Psir}
\Psi_{,r}(r,R)>0,\qquad\forall r>0,\quad\forall R\in [0,r].
\end{equation}

As mentioned above, we require the existence of a regular Cauchy
surface ($t=0$, say) carrying the initial data for the fields.
This requirement is fundamental, since it assures that the
singularities eventually forming will be a genuine outcome of the
dynamics. It is easy to show that, with the equation of state used
here, it is equivalent to require the matter density to be finite
and non vanishing on the data. Due to eqs. \eqref{eq:efe1b} and
\eqref{constraint} we get
\begin{equation}\label{pr2}
\lim_{r\to 0^+}\frac{\Psi_{,r}(r,r)}{r^2}\in (0,+\infty).
\end{equation}
Since area--radius coordinates map the whole set $\{(t,0):t\le
t_0\}$ into the point $R=r=0$, one may ask whether this may give
rise to some kind of contradiction, that is whether the
hypersurface $\{R=r\}$ fails to be regular.
However, note that the coordinate change, restricted on the
initial data hypersurface, is regular up to the centre, since
the generic point $(0,r)$ in comoving coordinates is mapped onto
the point $(r,r)$ in area--radius coordinates.
Moreover, we are going
to put analiticity of the data into play. In a neighboorood
of the center, this property has to be checked using a cartesian
system of coordinates, since even powers of $r$ can give rise to
loss of differentiability at finite order in such coordinates.

To inspect this point we consider the whole set of Cauchy data for
the fields. Let us choose a coordinate system on $\Sigma$ in such
a way that the embedding reads
\begin{equation}
\Sigma(\sigma,\theta,\phi)\hookrightarrow
M(r=\sigma,R=\sigma,\theta,\phi).
\end{equation}
The induced metric and the extrinsic curvature (i.e. the second
fundamental form) are respectively given by
\begin{align}
&\text ds^2_{\Sigma}=\frac{1}{4\pi E(\sigma)
\sigma^2\rho(\sigma,\sigma)}\text d\sigma^2+ \sigma^2\text
d\Omega^2,\\
&K_\Sigma=-\frac{u(\sigma,\sigma)}{8\pi
E(\sigma)}\left(\frac1{R^2\rho}\right)_{,R} (\sigma,\sigma)\,\text
d\sigma^2 - \sigma\,u(\sigma,\sigma) \text d\Omega^2.
\end{align}
It is now relatively easy to check that, if $\Psi(r,R)$ is
analytic and odd, and $Y_0(r)=1+O(r^2)$ is even,
using \eqref{eq:efe1b}, \eqref{eq:efe2b} and \eqref{eq:Y} the above tensors on
$\Sigma$ are analytic and even in $r$.

This means that all the
physical quantities give rise to analytic functions in cartesian
coordinates near the center.

Finally, we require regularity of the metric at the center that
is, in comoving coordinates:
\begin{equation}\label{eq:reg}
R(0,t)=0,\qquad e^{\lambda(0,t)}=R'(0,t),
\end{equation}
for each $t\ge 0$ up to the time of singularity formation $t_0$.

The singularity forms whenever the denominator in \eqref{eq:efe1b}
vanishes, that is $R=0$. This kind of singularity is called a {\sl
shell--focusing} singularity (we have excluded here, via
equations \eqref{eq:Rprime} and \eqref{eq:Psir} , the so called {\sl shell--crossing}
singularities at which the particle flow-lines intersect each
other). In comoving coordinates $(r,t)$, the locus of the zeroes
of $R(r,t)$ defines implicitly a singularity curve $t_s(r)$ via
$R(r,t_s(r))=0$. The quantity $t_s(r)$ represents the comoving
time at which the shell labeled $r$ becomes singular. The
singularity forms if $t_s(r)$ is finite for each shell. In
physically viable cases the curve $t_s(r)$ is strictly increasing
and the center is the first point which can become singular.
Regularity of the data then implies
\begin{equation}\label{eq:t0}
\lim_{r\to 0^+} t_s(r)=t_0>0.
\end{equation}

In order to describe the singuarity formation at the
shells $r>0$ by condition $R=0$, from \eqref{eq:efe1b}
we make the assumption
\begin{equation}
\lim_{R\to 0^+}\frac{\Psi_{,R}(r,R)}{R^2}=-\infty,
\end{equation}
for $r$ sufficiently close to 0. Using the above requirements,
toghether with \eqref{eq:Ydef}, we can also translate relations
\eqref{eq:reg} in area--radius coordinates asking
\begin{equation}\label{eq:Y0}
\lim_{r\to 0^+}Y(r,xr)=1,\qquad\forall x\in (0,1].
\end{equation}

\subsection{Taylor expansion of the mass}\label{subsec:ty}
As said in Section \ref{rem:rend}, in the present paper we assume
analyticity of the mass function at $(0,0)$.
It should be noticed that the `point` $(0,0)$
in mass-area coordinates `contains` both a regular part (it
contains the data $R=r$ as $r$ goes to zero) and a part at which
the spacetime becomes singular (as $R$ goes to zero along the
singularity curve, see next section). The mass function itself
however satisfies an equation which is regular at the spacetime
singularity, so that the assumption made here is
exactly equivalent to that usually made on the data in other
models of gravitational collapse. Such data can be taken to be
analytic in cartesian coordinates near the center, as in
\cite{e3}, or simply Taylor-expandable up to the required order as
in \cite{sj}). In the present paper however we assume analiticity.
Moreover, coherently
with our choice of initial data, we will assume odd--parity of the
mass function.

The following holds true:

\begin{prop}\label{prop:taylor}
The Taylor expansion of the mass function $\Psi(r,R)$ has the following
structure
\begin{equation}\label{eq:06}
\Psi(r,R)=\frac h2 \left(r^3 -\frac{\alpha}{\alpha+1}
R^3\right)+\sum_{i+j=3+k} \Psi_{ij}r^i R^j+\ldots.
\end{equation}
where $k$ is an even integer, $k\geq 2$ and $h$ is a positive constant.
\end{prop}

\begin{proof}
Odd parity of $\Psi$ and regularity condition \eqref{pr2} and
\eqref{constraint} imply
the Taylor expansion to
start from third order terms.
Therefore, one certainly has
\begin{equation}\label{eq:Psi-exp}
\Psi(r,R)=\sum_{i+j=3}\Psi_{ij}r^i
R^j+\ldots.
\end{equation}
For the sake of convenience we now set, for each $n\ge 0$,
\begin{equation}\label{eq:AkBk}
A_n(\tau)=\sum_{i+j=3+n}i\,\Psi_{ij}\tau^j,\qquad
B_n(\tau)=\sum_{i+j=3+n}j\,\Psi_{ij}\tau^{j-1},
\end{equation}
so that the $r^{n+2}$'s coefficients of Taylor expansions of
$\Psi_{,r}(r,r\tau)$ and $\Psi_{,R}(r,r\tau)$ are
$A_n(\tau)$ and $B_n(\tau)$ respectively.
We recall that \eqref{pr2} implies
$A_0(1)>0$, and, from \eqref{constraint}, $B_0(1)<0$ follows.
Using \eqref{eq:Ynew} we get
\[
Y(r,r\tau)= \frac{A_0(\tau)}{A_0(1)}
\left[\frac{B_0(1)\tau^2}{B_0(\tau)}\right]^\frac{\alpha}{\alpha+1}+o(1),
\]
at least for each $\tau\in(0,1]$ such that $B_0(\tau)\not=0$ (but
this polynomial can possibly vanish only for two values of $\tau$),
and then \eqref{eq:Y0} holds if
\begin{equation}\label{eq:necbound}
\frac{A_0(\tau)}{A_0(1)}
\left[\frac{B_0(1)\tau^2}{B_0(\tau)}\right]^\frac{\alpha}{\alpha+1}=
1,\qquad\forall\tau\in(0,1]\text{\ \ with $B_0(\tau)\not=0$}.
\end{equation}
But
\[
\frac{B_0(1)\tau^2}{B_0(\tau)}=\tau^2
\frac{\Psi_{12}+2\Psi_{21}+3\Psi_{03}}
{\Psi_{21}+2\Psi_{12}\tau+3\Psi_{03}\tau^2},
\]
and therefore if $\Psi_{21}$ was not vanishing, the above
quantity would tend to zero as $\tau\to 0$, which is in
contradiction with \eqref{eq:necbound}. Then $\Psi_{21}=0$. A
similar argument applies to $\Psi_{12}$ to show that this
quantity is zero as well. Finally, relation \eqref{constraint}
imposes a constraint on $A_n(1)$ and $B_n(1)$:
\begin{equation}\label{eq:constraint}
-\alpha\,A_n(1)=(\alpha+1)\,B_n(1),\qquad\forall n\ge 0.
\end{equation}
Using this equation for $n=0$ and setting $h:=2A_0(1)$ we finally
get formula \eqref{eq:06}.
\end{proof}

\begin{rem}\label{rem:compat}
A tedious but straightforward calculation shows that the Taylor
expansion \eqref{eq:06} is compatible with \eqref{eq:2PDE} "in
the Cauchy-Kowaleski sense" at {\it any} order, that is, the
equation allows the iterative calculation of all the higher order
terms once the data are chosen. Of course, we stress that this is
{\it not} a proof of global existence up to singularity formation
 but only a -
fundamental - consistency check for solutions here assumed a
priori as regular.
\end{rem}

\begin{rem}\label{rem:co} The Taylor expansion given above excludes the
self-similar solutions from what follows. It can, in fact, be
easily shown that analyticity in self-similar variables leads to
mass functions of the form $\Psi=r\tilde\psi(R/r)$ where
$\tilde\Psi$ is finite at $R=0$. One recovers here a fact which is
very well known in the case of dust spacetimes, where the
self-similar mass profile has a constant $\tilde\Psi$ (linear
profile) while analyticity of the data for non self-similar
solutions requires $\Psi$ to start from cubic terms.
\end{rem}

\subsection{The apparent horizon}
A key role in the study of the nature of a singularity is played
by the apparent horizon $t_h(r)$ defined in \eqref{eq:horizon}
(see for instance \cite{jbook}). The apparent horizon is the
boundary of the trapped surfaces, and therefore represents the
comoving time at which the shell labeled $r$ becomes trapped. In
area-radius coordinates this boundary is defined by
$R_{hor}=2\Psi(r,R_{hor})$. Since $\Psi_{,R}(0,0)=0$, implicit function
theorem ensures that the curve $R_{hor}$ is defined in a right
neighborhood of $r=0$. In what follows, we shall need the
behavior of this curve near $r=0$. It is easy to check that $R_{hor}$
is strictly increasing and such that $R_{hor}(r)<r$. Moreover it is
$R_{hor}(r)\cong 2\Psi(r,0)$, since from \eqref{eq:horizon} it is
\[
R_{hor}=2\Psi(r,0)+2 R_{hor} \Psi_{,R}(r,0)+R_{hor}^2\, g(r,R_{hor}),
\]
where $g$ is bounded and $\Psi_{,R}(r,0)$ is infinitesimal.
Therefore, due to eq. \eqref{eq:06}, we conclude that
\begin{equation}\label{eq:hor}
R_{h}(r)=h r^{3}+\ldots.
\end{equation}

Next section is devoted to the study of the nature of the central
($R=r=0$) singularity. We restrict ourselves to this singularity
since, in barotropic perfect fluid models with positive pressures,
it is the only one that can be naked. This is easily seen using
comoving coordinates. Indeed, a singularity cannot be naked if it
occurs after the formation of the apparent horizon (i.e. it must
be $t_{h}(r)\geq t_s(r)$). A necessary condition for this is that
the singularity must be massless ($\Psi(r,t_s(r))=0$). But, due to
equation \eqref{ma2}, in presence of a positive pressure the mass is
strictly increasing in a collapsing ($\dot R<0$) situation, while
it is zero at the regular centre. The situation can be completely
different if negative pressures are allowed: in this case non
central singularities can be naked as well \cite{cjj}.

\subsection{Nakedness of the central singularity}

At the center ($R=r=0$) the apparent horizon and the singularity
form simultaneously and the necessary condition for nakedness is
satisfied. The singularity will be (locally) naked if there exists
a radial lightlike future pointing local solution $R_g(r)$ of the
geodesic equation with initial condition $R(0)=0$ "travelling
before the apparent horizon", that is - in area radius
coordinates - $R_g(r)>R_{hor}(r)$ for $r>0$. We will study in full
details only the existence of {\it radial} null geodesics
emanating from the singularity. It can in fact be proved that, if
a singularity is radially censored (that is, no radial null
geodesics escape), then it is censored \cite{Nolan,ns}.

The equation of radial null geodesics in the coordinate system
$(r,R)$ is easily found from \eqref{met} setting $\text ds^2=0$
together with $\text d\theta=\text d\phi=0$:
\begin{equation}\label{eq:geo-short}
\frac{\text dR}{\text dr}=
-\frac{\alpha}{\alpha+1}\frac{\Psi_{,r}}{\Psi_{,R}}
\left(1-\frac{u}{Y}\right).
\end{equation}
Our main result can be stated as follows:

\begin{teo}\label{naked}
For any choice of initial data $Y_0(r)$, $\Psi_0(r)$ for the
Einstein field equations such that
\begin{enumerate}

\item\label{uno}  the central singularity forms in a
finite amount of comoving time, and
\item the Taylor expansion of the mass function is given by \eqref{eq:06},
\end{enumerate}
there exists solutions of \eqref{eq:geo-short} that extend back to the
central singularity, which is therefore locally naked.
\end{teo}

To show the result we first need the following lemma.

\begin{lem}\label{lem:tx}
Called $t_x(r)$ the curve defined by $R(r,t_x(r))=x r^3$, there
exists a $x>h$ such that
\begin{equation}\label{eq:control}
\lim_{r\to 0^+}t_x(r)=t_0.
\end{equation}
\end{lem}

\begin{proof}
It must be shown that for some $x>h$
\begin{equation}\label{eq:tx}
\lim_{r\to 0^+}\int_0^{x
r^3}\frac{\rho^\alpha(r,\sigma)}{u(r,\sigma)}\,\text d\sigma=0.
\end{equation}
With the variable change $\sigma=\tau r^3$ the integral above
becomes
\[
r^3\int_0^x \frac{\rho^\alpha(r,r^3\tau) r^{3/2}\sqrt\tau}
{\left(2\Psi(r,r^3\tau)+\tau r^3(Y^2(r,r^3\tau)-1)\right)^{1/2}}\,\text
d\tau,
\]
and to prove \eqref{eq:tx}
using Fatou's lemma it suffices to show that
\begin{equation}\label{eq:fatou}
\int_0^x \lim\sup_{r\to 0^+}\left(\frac{\rho^\alpha(r,r^3\tau) \sqrt\tau}
{\left[\left(\frac{2\Psi(r,r^3\tau)}{r^3}-\tau\right)+\tau\,Y^2(r,r^3\tau)\right]^{1/2}}\right)\,\text
d\tau<+\infty.
\end{equation}
We first notice that the quantity in square
brackets at the denominator in the above expression must be
positive for $r$ small. This is to ensure dynamics near the
central singularity (see, e.g., \eqref{eq:efe2b}). But, using \eqref{eq:06},
it is
\[
\left(\frac{2\Psi(r,r^3\tau)}{r^3}-\tau\right)=(h-\tau)+O(r^2),
\]
where $O(r^2)$ is infinitesimal uniformly in $\tau$ (again,
this notation  means infinitesimal
behaviour, uniform in $\tau$). Since $\tau$ can be greater than $h$, then
$Y(r,r^3\tau)$ cannot be infinitesimal as $r$ goes to 0.
Recalling $Y=\frac{E(r)\Psi_{,r}}{(\alpha+1)\rho^\alpha}$,
and exploiting \eqref{eq:Y0} for $x=1$, it is also a simple
task to check that $E(r)$ behaves like $r^{-2}$,
\[
E(r)\,\Psi_{,r}(r,r^3\tau)=c_0+O(r),
\]
and so $\rho^{\alpha}(r,r^3\tau)$ cannot be infinite as $r$
approaches 0.
The expression for $\rho$ is given by
\eqref{eq:efe1b}; for simplicity we compute $\rho^{\alpha+1}$, using
\eqref{eq:06}:
\begin{multline*}
\rho^{\alpha+1}(r,r^3\tau)=-c_1\frac{\Psi_{,R}(r,r^3\tau)}{r^6\tau^2}
=c_1\cdot \\
\cdot\left[\frac32h\frac\alpha{\alpha+1}-\frac1{\tau^2}
\left(\frac{\Psi_{41}}{r^2}+\Psi_{61}+O(r)
\right)-\frac2\tau\left(\Psi_{32}+O(r)\right)+O(r)\right],
\end{multline*}
where $c_1$ is the positive constant $(4\pi\alpha)^{-1}$. As said
above, this cannot be infinite and therefore $\Psi_{41}$
vanishes, giving for some constant $c_2$
\[
\rho^{\alpha}(r,r^3\tau)=\frac{c_2}{\tau^\frac{2\alpha}{\alpha+1}}
(b(\tau)+O(r))^\frac{\alpha}{\alpha+1},
\]
where $b(\tau)$ is a regular function. This yields, passing to the
limit $r\to 0^+$, the following expression for the integral in
\eqref{eq:fatou}:
\begin{equation}\label{eq:limsup}
\int_0^x\frac{c_2 b(\tau)^\frac{2\alpha}{\alpha+1}\sqrt\tau}
{\tau^\frac{2\alpha}{\alpha+1}\left[(h-\tau)b(\tau)^\frac{2\alpha}{\alpha+1}+\tau
c_0^2 c_2^2 \tau^\frac{4\alpha}{\alpha+1}\right]^{1/2}}\,\text
d\tau.
\end{equation}
The term in square bracket at the denominator is bounded away from zero for
$\tau\le h$  and so is for $x$ greater than but sufficiently near to $h$.
Recalling the bound $\alpha<1$, the above integral is therefore finite, and
the lemma is proved.

\end{proof}

\begin{rem}

Let us observe that we have incidentally shown here that
\begin{equation}\label{eq:Rx}
\Psi_{,R}(r,x r^3)=- a(x) r^6+\ldots,\quad\left(1-\frac uY\right)(r,x r^3)=d(x)+\ldots,
\end{equation}
where $a(x)$ and $d(x)$ are some positive functions.

Also observe that the same argument of the above lemma can be used to show
that also $t_{hor}(r)$ tends to $t_0$ as $r\to 0^+$, that is the
centre gets trapped at the same comoving time it becomes singular.
\end{rem}

\begin{proof}[Proof of theorem \ref{naked}]
To show the existence of singular geodesics we use a simple
technique developed earlier \cite{ns}. First of all, we recall
that a function $y_0(r)$ is called a subsolution (respectively
supersolution) of an ordinary differential equation of the kind
$y'=f(r,y)$ if it satisfies $y_0'\leq f(r,y_0)$ (respectively
$\geq$). Now, it can be shown \cite{GGM} that the apparent horizon
$R_{h}(r)$ is a supersolution of the geodesic equation
\eqref{eq:geo-short}. The singularity is certainly naked if it is
possible to find a subsolution $R_+(r)$ of the same equation which
stays over the horizon. In fact, choose a point $(r_0,R_0)$ in
the region $S =\{(r,R)\,:\, r>0, R_{hor}(r)<R<R_+(r)\}$. At this
point the (regular) Cauchy problem with datum $R(r_0)=R_0$ admits
a unique local solution $R_g(r)$. Now the extension of this
solution in the past cannot escape from $S$ since either it would
cross the supersolution from above or it would cross the
subsolution from below. Thus it must extend back to the
singularity with $\lim_{r\to 0^+}R_g(r) =0$.

We now proceed to show that a subsolution always exist. For this
aim, it suffices to consider a curve $R_x(r)=x r^3$, with $x>h$.
Indeed, computing the righthand side of \eqref{eq:geo-short} for
$R_x(r)$, using \eqref{eq:Rx}, we get that $R_x(r)$ is certainly a subsolution of
\eqref{eq:geo-short} if
\begin{equation}\label{eq:subs2}
x <\frac\alpha{\alpha+1}\,\frac{h}{2 a(x) r^4} d(x),
\end{equation}
that is always satisfied, independently of $x$, for $r$ sufficiently small.

Therefore, if we consider the curve $R_x(r)$ for $x>h$ sufficiently
near to $h$, then Lemma \ref{lem:tx} ensures that --
re--translated in comoving coordinates -- it emanates
from the central singularity, and so the theorem is proved.

\end{proof}

We stress that the theorem holds for any solution satisfiyng
\eqref{eq:t0} and \eqref{eq:06}.

\section{Extension to the general barotropic case}

We are going to show in the present section that our main result,
namely the existence of naked singularities,
actually hold for the general (i.e. not necessarily linear) barotropic equation of state
$\epsilon=\epsilon(\rho)$ provided that a set of
(physically motivated) requirements are satisfied
by the state function:
\begin{assum}\label{assum:gen}
We assume $\epsilon=\epsilon(\rho)$ to be a $\mathcal C^1$ function
in $[\bar\rho,+\infty)$ (where $\bar\rho\ge 0$), such that
$\epsilon(\rho)\ge 0\,(=0\text{\ iff\ }\rho=\bar\rho)$.
Recalling \eqref{pe}, that is
$p(\rho)= \rho\frac{\text d\epsilon}{\text d\rho}-\epsilon
$, we also assume $p(\rho)$ is a strictly positive $\mathcal
C^1$ function with $\frac{\text dp}{\text d\rho}>0$,
except at most for a bounded interval $[\bar\rho,\rho_1]$,
possibly coinciding with a single point,
where $p(\rho)$ can vanish.
\end{assum}

\begin{rem}\label{rem:gen}
Observe that:
\begin{enumerate}
\item\label{rem:p1} The assumptions made imply that
\begin{equation}\label{eq:eprime}
\frac{\text d\epsilon}{\text d\rho}(\rho)>0 \text{\ if\ } \rho>\bar\rho.
\end{equation}
and therefore $\epsilon(\rho)$ is a strictly increasing positive
function.
\item Differentiating \eqref{pe} we have, where it makes sense,
\begin{equation}\label{eq:state-diff}
\frac{\text dp}{\text d\rho}=\rho\,\frac{\text d^2\epsilon}{\text
d\rho^2},
\end{equation}
then $\epsilon(\rho)$ is strictly convex
for $\rho$ sufficiently large, and so
\begin{equation}\label{eq:assum2}
\lim_{\rho\to +\infty} \frac{\text d\epsilon}{\text
d\rho}(\rho)=+\infty.
\end{equation}
\item
The assumptions made imply the existence of
$\lim_{\rho\to\infty}p(\rho)$. In addition, if the limit would be finite, say $l$,
then we should have $\frac{\text d\epsilon}{\text
d\rho}(\rho)<\frac1\rho(l+\epsilon(\rho))$, and then
$\epsilon(\rho)<\rho+l$ by a simple comparison argument in
o.d.e., which is in contradiction with \eqref{eq:assum2}. Thus
\begin{equation}\label{eq:p-gen}
\lim_{\rho\to \infty}p(\rho)=+\infty.
\end{equation}
\end{enumerate}
\end{rem}

\begin{rem}
We stress that the above mentioned hypotheses are quite natural
from the physical point of view. Besides obviously including the
$p=\alpha \epsilon$ equation of state considered so far, they
include, for instance, the equation of state of the perfect gas
$p(\rho)=K_2\rho$ for which
$\epsilon(\rho)=K_1\rho+K_2\rho\log\rho$ where $K_1 $ and $K_2$
are positive constants (in this case one obviously has
$\rho_1=\bar\rho=e^{-{K_1}/{K_2}}$).
\end{rem}

Einstein's equation \eqref{ma2} reads
\begin{equation}\label{eq:press-gen}
p=-\frac{\Psi_{,R}}{4\pi R^2}.
\end{equation}
Using it, together with \eqref{ma1},
\eqref{pe}, \eqref{eq:rho} and the coordinate change formulae
$\Psi'=\Psi_{,r}+R'\,\Psi_{,R}$ and $\dot\Psi=\dot R\,\Psi_{,R}$,
we obtain the general counterparts for equations \eqref{eq:Rprime} and
\eqref{eq:Y}, namely
\begin{equation}\label{eq:Rprime-gen}
R'=-\frac{\Psi_{,r}}{\Psi_{,R}}\,\frac{p}{\epsilon+p}=
\frac{\Psi_{,r}}{4\pi R^2 (\epsilon+p)},
\end{equation}
and
\begin{equation}\label{eq:Y-gen}
Y(r,R)=\frac{E(r)\Psi_{,r}(r,R)}{(\text d\epsilon/\text
d\rho)(\rho(r,R))}.
\end{equation}
Using these formulae, we can again express the metric in the form
\eqref{met}. The crucial point is now that we can express
$\frac{\text d\epsilon}{\text d\rho}(\rho)$ as a function of
$p(\rho)$ and, as a consequence, the dynamics of the system is,
also in the general case, expressed in terms of the mass function
and its derivatives only. To this end, consider the parameterized
curve in $\Re^2$
\[
\Re^+\ni\rho\mapsto
\left(\xi(\rho)=p(\rho),\zeta(\rho)=\frac{\text d\epsilon}{\text
d\rho}(\rho)\right).
\]
This curve is globally the graphic of a function
$\zeta=\zeta(\xi)$, recalling that $\xi$ is a non-decreasing
function of $\rho$ (by the assumptions on $p(\rho)$), and that, by
\eqref{eq:state-diff}, $\frac{(\text d\zeta/\text d\rho)}{(\text
d\xi/\text d\rho)}=\rho\not=0$. Since also $\zeta$ is increasing
by \eqref{rem:p1} of Remark \ref{rem:gen}, there exists
$\lim_{\rho\to \bar\rho^+}\zeta(\rho)=\zeta_0$ finite. Denoting by
$\xi_0$ the finite number $\lim_{\rho\to\bar\rho^+}\xi(\rho)$, the
function may be prolonged up to the point $(\xi_0,\zeta_0)$. Let
also observe that this function is $\mathcal C^1$, for each
$\xi>\xi_0$, where indeed $\frac{\text d\xi(\rho)}{\text d\rho}$
is strictly positive by the assumptions made on $p(\rho)$.

Using this result, and recalling
\eqref{pe}, one finds that $Y$ in \eqref{eq:Y-gen} (and
then $u$ in \eqref{eq:efe2b}) can be expressed as functions of the
data and of the mass function $\Psi(r,R)$ and its derivatives.
Then, again, with some calculations one obtains a second order PDE
that must be satisfied by $\Psi$. As in the case treated so far,
we consider only analytic solutions of this equation, and
proceed to analyze the structure of the lower order terms of the
mass profile.

First of all, since $R'\equiv 1$ on the data surface $R=r$, the
expression for the initial energy is:
\[
\epsilon(r,r)=\frac{\Psi_{,r}(r,r)+\Psi_{,R}(r,r)}{4\pi r^2}.
\]
Imposing the regularity condition $\lim_{r\to
0^+}\epsilon(r,r)\in(0,+\infty)$ and making reference to the
notation used in Section \ref{subsec:ty} we get
\[
A_0(1)+B_0(1)>0.
\]
Actually, $A_0(1)\not=0$. Otherwise, $\epsilon(r,r)+p(r,r)\to 0$
as $r\to 0^+$, since
$\epsilon(r,r)+p(r,r)=\frac{\Psi_{,r}(r,r)}{4\pi r^2}=
\frac{A_0(1)}{4\pi}+o(1)$. But $(\epsilon+p)(\rho)$ is a strictly
increasing and non negative function of $\rho$,
then it would be $\rho(r,r)\to\bar\rho$,
which would imply $\epsilon(r,r)\to 0$, that is a contradiction.

As in section \ref{subsec:phys}, for physical reasonableness we suppose the
initial energy $\epsilon(r,r)$ (and therefore $\rho(r,r)$) to be a
non increasing function of $r$. This implies that we can consider,
without loss of generality, the case in which also $B_0(1)\not=0$.
In fact, if $B_0(1)$ vanishes by \eqref{pe} it has to be
$p(r,r)\to 0$ as $r\to 0^+$. This fact, recalling the assumptions made on
the pressure, shows that $p(r,r)$ (that
is a non increasing function of $r$) must be identically zero. But
$\rho$ (and therefore $p$) must diverge at the spacetime
singularity, and therefore there exists an hypersurface, such that
$p$ is non zero but the energy $\epsilon$ is still regular, where
we can re--assign the initial data on. On this hypersurface, the
pressure must converge to a finite non--zero value as $r\to 0^+$.
Then we will suppose $B_0(1)\not=0$.
Finally, we note that positivity of pressure on the data further
implies that $A_0(1)>0$ and $B_0(1)<0$.

We are now ready to investigate lower order terms in the mass
function. Recall that regularity of pressure along the initial data implies
that $\Psi$ cannot contain first order terms (see
\eqref{eq:press-gen}). Then, as in \eqref{eq:Psi-exp} we set
\[
\Psi(r,R)=\sum_{i+j=3}\Psi_{ij}r^i
R^j+\ldots.
\]
We now denote by $\epsilon_0>0$ the limit $\lim_{r\to
0^+}\epsilon(r,r)$. By \eqref{rem:p1} of Remark \ref{rem:gen},
there exists a unique $\rho_0>0$ such that
$\epsilon(\rho_0)=\epsilon_0$, and clearly $\rho_0=\lim_{r\to
0}\rho(r,r)$. We also denote by $\beta_0$ the positive number
$\frac{\text d\epsilon}{\text d\rho}(\rho_0)$. Using \eqref{eq:Y0}
at $\tau=1$ we have
\begin{equation}\label{eq:E-gen}
E(r)=\frac{\beta_0}{A_0(1)}\,\frac 1{r^2}+\ldots,
\end{equation}
plus higher order terms.
Observe now that, for a fixed $\tau$,
\[
p(r,r\tau)=\frac{B_0(\tau)}{4\pi\tau^2}+\ldots=
-\frac{1}{4\pi}\left(\frac{\Psi_{21}}{\tau^2}+
2\frac{\Psi_{12}}\tau+3\Psi_{03}\right)\equiv p_0(\tau).
\]
If $(\Psi_{21},\Psi_{12})\not=(0,0)$, then $p_0(\tau)\to\infty$ as
$\tau\to 0$, and so $\rho(r,r\tau)$ (and therefore $\frac{\text
d\epsilon}{\text d\rho} (\rho(r,r\tau))$) is sufficiently large,
for $\tau$ near to $0$.
This leads to a contradiction, since using
\eqref{eq:E-gen} in \eqref{eq:Y-gen} shows that
$Y(r,r\tau)\cong\frac{\mu}{(\text d\epsilon/\text d\rho)(r,r\tau)}$,
for some non--zero constant $\mu$ independent of $\tau$,
but \eqref{eq:Y0} must hold.
Then, again, $B_0(\tau)=B_0(1)\tau^2$, and
$p_0(\tau)=p_0(1)\equiv p_0>0$.

Then the above argument shows that the lower order
terms of the mass have the structure, analogue to
\eqref{eq:06},
\begin{equation}\label{eqn}
\Psi(r,R)=\frac h2 \left(r^3 -\frac{p_0}{\epsilon_0+p_0} R^3\right)+\ldots.
\end{equation}
We now proceed analyzing the nature of the singularity forming at
the center.
With arguments similar to Lemma \ref{lem:tx}, opportunely
modified, it can be checked that some of the curves $R_x=x\,r^3$,
for $x>h$ sufficiently near, are emanating from the central
singularity (if seen in comoving coordinates). Indeed, we first
observe that, in the case of a barotropic equation of state,
\eqref{ma4} yields $-\text d\nu=\frac{\text dp}{\epsilon+p}\le
\frac{\text dp}{2p}$, where the inequality is given by dominant
energy condition $\epsilon-p\ge 0$. This implies
$e^{-\nu}\le\sqrt{p}$, and so the counterpart for the integral in
\eqref{eq:fatou} in this case has the following upper bound
\begin{multline*}
\int_0^x \lim\sup_{r\to 0^+}\left(\frac{e^{-\nu}(r,r^3\tau) \sqrt\tau}
{\left[\left(\frac{2\Psi(r,r^3\tau)}{r^3}-\tau\right)+\tau\,Y^2(r,r^3\tau)\right]^{1/2}}\right)\,\text
d\tau\le\\ \le
\int_0^x \lim\sup_{r\to 0^+}\left(\frac{\sqrt{p(r,r^3\tau)\tau}}
{\left[\left(\frac{2\Psi(r,r^3\tau)}{r^3}-\tau\right)+\tau\,Y^2(r,r^3\tau)\right]^{1/2}}\right)\,\text
d\tau.
\end{multline*}
Taking into account \eqref{eq:Rx} (that still holds) in \eqref{eq:press-gen}
to evaluate $p(r,r^3\tau)$, one can see that the integral above
takes a similar form to \eqref{eq:limsup}:
\[
\int_0^x\frac{c\,b_1(\tau)}
{\sqrt\tau\left[(h-\tau)b_1(\tau)+\tau b_2(\tau)\right]^{1/2}}\,\text
d\tau.
\]
and so it is finite, as in Lemma \ref{lem:tx}.

Now, using \eqref{eq:Rprime-gen}, one can compute both
sides of the
null radial geodesic equation
\begin{equation}\label{eq:geo-shortn}
\frac{\text dR}{\text dr}= R' \left(1-\frac{u}{Y}\right)
\end{equation}
for $R=x r^3$, obtaining a similar expression to
\eqref{eq:subs2}. We only remark that in this case, since $Y(r,r^3\tau)$ cannot be
infinitesimal as $r$ goes to zero, then $\frac{\text
d\epsilon}{\text d\rho}(\rho(r,r^3\tau)$ is finite (see
\eqref{eq:Y-gen}) and so is $\rho(r,r^3\tau)$.
We can therefore conclude
this section with the analogue of Theorem \ref{naked}, that is:
\begin{teo}\label{naked-gen}
Under the hypotheses made on the equation of state in
the assumption \ref{assum:gen}, for any choice of initial data for the
Einstein field equations such that
\begin{enumerate}
\item  the central singularity forms in a
finite amount of comoving time, and
\item the Taylor expansion of the mass function is given by \eqref{eqn},
\end{enumerate}
there exists solutions of
\eqref{eq:geo-shortn} that extend back to the central singularity,
which is therefore locally naked.
\end{teo}

\section{Discussion and conclusions}

Up today all analytical studies on naked singularities formation
in collapsing matter of astrophysical interest (i.e. fluids) have
assumed simplyfing hipotheses such as dust or self-similarity.

We have shown here for the first time that among non self-similar
barotropic perfect fluid solutions, all those describing complete
collapse for which the mass function is regular in a neighborhood
of the regular center up to singularity formation form naked
singularities. Besides of spherical symmetry, this result is
independent on any simplifying assumption.

The problem of the classification of the data which leads to such
singularities remains for future work. In particular, it is
unclear if the set generating naked singularities is really of
non-zero measure in the space of the data or not.

\end{document}